# Puzzlegram: a Serious Game Designed for the Elderly in Group Settings

Sunny Choi [0000-0003-3823-1149]

MusEdLab, New York University, New York NY 10012, USA
`ssc526@nyu.edu`

**Abstract.** An original serious game prototype named 'Puzzlegram' is created for the elderly demographic in group settings as the target players. Puzzlegram is precisely designed to accentuate memory, auditory interaction as well as haptic response to visual signals with the use of music. Music is introduced as a key component for establishing the game design that provides a source of meaningful contextualization — familiar music from the past — for setting the game mechanics, which facilitated the construction of the serious game design process. The discussion topics raised include the need to design serious games for fostering meaningful interactions, as well as developing a thorough framework for constructing purposeful design for serious games. A potential integral of artificial intelligence to Puzzlegram may involve assigning a novel dimension to its existing problem-solving task by adapting to varying states of cognitive function for monitoring purposes based on an individual's interaction with the game.

## 1   Introduction

While different games serve different purposes, video games are widely perceived as casual activities with entertainment value [1]. Serious games, on the other hand, are positioned with a goal to deliver beyond entertainment purposes, such as developing new skills, conveying meaning, and providing experiences and emotions or changing behavior and attitudes [2]. Just as general video games and serious games have depth-contrasting goals set to be met by the players, game design techniques can also be initially established with a goal to fulfill specific commercialization or academic/professional purposes [3].

When it comes to applying characteristics of serious games for a specific age group such as the elderly demographic, studies have said to primarily consider various physical and cognitive training to improve elderly people's deteriorating physiological functions [4]. Deteriorating bodily functions are evidently not conditions to be entertained with, thus creating a serious game for the elderly demographic requires a thoughtfully constructed game design process that can genuinely evoke a deeper purpose for the elderly people's fundamental well-being. For instance, obtaining a higher score in a generic video game played by an elderly player does not necessarily reflect a healthy cognitive state of the player. Despite a wide range of new products introduced as serious games to serve the elderly population, the issue is that such products have already been classified as regular games with no specific aim to serve the elderly, lacking originality and rigorous game design research by leaving out the very target group that the product apparently claims to serve [4].

In this concisely formatted paper, the author introduces 'Puzzlegram': an original serious game prototype created for the elderly players in group settings as the target group. The game is specifically designed to accentuate memory, auditory interaction as well as haptic response to visual signals with the use of music as an essential component of the game design.

## 2      Methodology

The underpinning theoretical framework for Puzzlegram's design as a serious game stems from the capability approach [5], which highlights a person's capability to function, where a sense of functioning specifically refers to what the person can do or can be from recognizing the non-materialistic value from effective opportunities [6]. Previous study [7] discusses socially responsible design as a capability approach-driven design for societal development that takes human diversity into account. By deeply considering the moral values of elderly people towards designing for society, capability approach leads to capability expansion, promoting advancement in design for collaborative development between designers, engineers, sociologists and scientists. A shared cultural commonality, such as autobiographical memory of music, is said to promote anticipation and expectation – two key factors that drive motivation which have been described as fundamental for the musical experience to understand the effects of music on emotion [8] – towards increasing engagement with the game interface. In other words, familiar music from the past may play an impactful role in facilitating a purposeful game design to evoke meaningful interactions.

## 3      Design

### 3.1      Game Overview

Puzzlegram is built as a software-led hardware serious game where music plays an essential role in establishing the game design mechanics to yield meaningful interactions through gameplay [9]. As a three-player game, each player is individually assigned his/her own Puzzlegram hardware controller to elicit a sense of ownership and responsibility as a unique contributor:

**Goal**. The goal of Puzzlegram is for all players to reach the end of the game as a team. By design as a non-competitive game, it is not possible for any one player to 'win' over other players (no ranking). Upon reaching the end of the game, players would have collaboratively constructed a familiar song from start to finish.

**Scope**. Puzzlegram is intentionally designed to foster collaborative learning and design research process by directly involving the players and incorporating their interaction with the game mechanics for further design analysis [10]. The credibility of Puzzlegram's design is therefore to be determined by the players' subjective degree of capability as individuals derived by the embedded game mechanics.

**Game mode**. Puzzlegram as a serious game is to be played with sound 'on', as sound plays an essential role in forming the game mechanics. If desired, the game may also be played muted, which would simply convert the game type as a memory game.

**Rule**. Upon pairing the three individual hardware pieces connected via Bluetooth, Puzzlegram officially begins when a short musical excerpt from a well-known tune is played represented by a unique solid color display. The game interface is displayed on a separate monitor, which also shows the user's haptic input response with a corresponding color that is assigned to each haptic region as output, which is entirely initiated from each player's own Puzzlegram hardware controller. Once the anticipated color is discovered and matched by all players, the game proceeds to the next segment by unveiling the next excerpt of the same musical tune until all players collaboratively reach constructing the entire musical tune from start to finish (Figure 1, 2).

**Tutorial**. Intentionally, there is no formal tutorial provided that explains the game rules. Instead, players are actively encouraged to explore the game hardware by freely touching any 16 unique haptic regions as the game is in active session. While there is only one haptic region that correctly corresponds to a matching reference color per 'level', there is no penalty associated with the user's game behavior for pressing any of the haptic regions where the corresponding color does not match the reference color.

**Level/progress**. The game does not increase or decrease in difficulty as it progresses. However, prior exposure to the color assigned to each haptic region as the game progresses may potentially make it easier to discover the matching color in a shorter time period.

**Challenge**. There are no additional challenge or obstacle components added to the game design.

**Reward**. When all three players press the expected haptic region that unveils the matching color display, the next musical excerpt is 'unlocked' and played in a loop, coupled with a new solid color display. As the game progresses, players reach closer to constructing the full song.

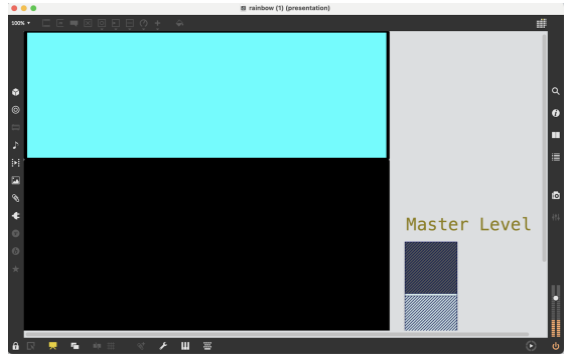

**Fig. 1.** Default visual display of Puzzlegram. The top rectangular solid color displays the corresponding color to be individually located and matched by each player.

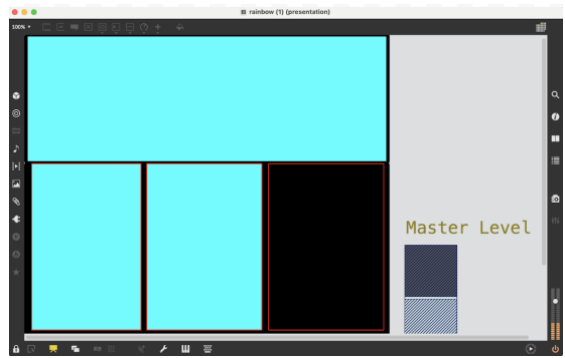

**Fig. 2.** User interface of Puzzlegram with two players waiting for the third player to locate the matching color.

### 3.2   Music and Game Mechanics

Several digital audio station software (DAW) programs were used to design the sound component for Puzzlegram. A popular song title that is instantly recognizable among elderly people in North America was selected as the prototype song. The song was re-arranged and composed with new instrumentation layers using Logic Pro (https://www.apple.com/logic-pro/). The creation of four distinct instrumentation layers composed of melody, harmony 1, harmony 2, and harmony 3. Using Ableton (https://www.ableton.com/), the exported audio layers were equally split to create a total of 16 distinct musical segments per instrumentation layer. Each audio file was randomly assigned an order between 1 and 16 and was paired to a haptic region.

Incorporating music to establish the game mechanics served an essential role in constructing Puzzlegram as a serious game. Multiple key components attached to music such as nostalgia, familiarity, and relatability have been considered to foster player motivation and add purpose to the serious game design. While it is possible to play Puzzlegram in a muted state with no music heard in the background, Puzzlegram would simply become a memory game to pair matching colors. In other words, making modifications to the original game may alter the function as a serious. Previous study also states that sound design can have a significant impact on user experience and attention in serious games [11].

### 3.3    Visual Design

Identically shaped white square areas are each assigned to the 16 haptic regions to distribute equal significance to all haptic regions on the hardware (Figure 3). By assigning equal shape and color to the interactable regions on the hardware, the emphasis is placed on the player's interaction with the controller rather than on the design which may influence the player behavior during the game. Puzzlegram's controller as hardware serves no function on its own, unless the player engages with it by interacting with any of the 16 haptic regions. Each haptic region is associated with a specific solid color display, visible from a separate monitor that displays the game interface. A randomly assigned music segment from an assigned instrumentation layer is also associated with each haptic region.

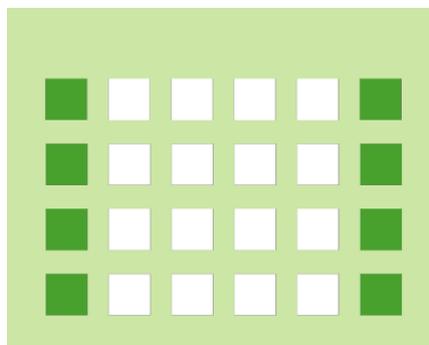

**Fig. 3.** Visual design for the hardware controller of Puzzlegram

## 4      Discussion Topics for Future Research

### 4.1      Serious Game for Fostering Meaningful Interactions

The key elements used in establishing the game mechanics for Puzzlegram stemmed from incorporating a proximate source of familiarity through music, which acted as a gateway for players to have a reason to participate in the game. The game design for Puzzlegram also explored placing priority on the player's degree of freedom while keeping the game engaging and entertaining, rather than instructing to master the game environment as an end-goal in itself.

### 4.2      Framework for Designing Purposeful Serious Games

The digital interface for Puzzlegram was entirely empowered by each individual player's haptic input actions. Rather than letting the game rules or the visual environment dictate user's gameplay behavior, a critical need to consider the process of serious game design with the demographic it intends to serve is essential. To develop an effective serious game from its conception to execution phase, a clear need to establish a framework for designing purposeful serious games is crucial to justify its differentiation from regular games.

## 5      Conclusion

This paper described the design of Puzzlegram as a serious game. The game design attempted to highlight the value behind its purposeful design which fosters individual capability for the players, particularly for the elderly people in group settings. A potential integral of artificial intelligence to Puzzlegram may involve assigning a novel dimension to its existing problem-solving task by adapting to varying states of cognitive function for monitoring purposes based on an individual's interaction with the game. In conclusion, Puzzlegram sparks an important question about the need to create serious games that serve the elderly that promote a sense of empowerment by design, rather than positioning them as simply recipients of care or need. Reaching beyond the subject of quality of design, which is ultimately a subjective matter, this paper hopes to initiate further research work towards what it means to serve and aid the globally aging society.